\documentclass[pre, preprint, amssymb]{revtex4}

\usepackage[pdftex]{graphicx}
\usepackage{bm}          
\usepackage{amsmath}

\newcommand{\REM}[1]{}
\newcommand{\gmag}{\Gamma}
\newcommand{\gvec}{\bm{\Gamma}}

\newcommand{\photovolt}{\mathcal{E}}

\newcommand{\fig}[1]{Fig.~\ref{#1}}
\newcommand{\eqn}[1]{Eqn.~\ref{#1}}
\newcommand{\Fig}[1]{Fig.~\ref{#1}}

\newcommand{\sct}[1]{Sec.~\ref{#1}}
\newcommand{\ten}[1]{\ensuremath{\times 10^{#1}}}
\newcommand{\unit}[1]{\ensuremath{\mathrm{\: #1}}}

\newcommand{\OhnesorgeNumber}{\ensuremath{\operatorname{Oh}}}

\begin{document}

\title{Vibrations of a diamagnetically levitated water droplet}

\author{R.J.A. Hill}
\author{L. Eaves}
\affiliation{School of Physics and Astronomy, University of Nottingham, Nottingham NG7 2RD, UK}
\email{richard.hill@nottingham.ac.uk}

\date{\today}

\begin{abstract}
We measure the frequencies of small-amplitude shape oscillations of a magnetically-levitated water droplet. The droplet levitates in a magnetogravitational potential trap. The restoring forces of the trap, acting on the droplet's surface in addition to the surface tension, increase the frequency of the oscillations. We derive the eigenfrequencies of the normal mode vibrations of a spherical droplet in the trap and compare them with our experimental measurements. We also consider the effect of the shape of the potential trap on the eigenfrequencies.
\end{abstract}


\maketitle
\section{Introduction}
If the surface of a spherical liquid drop is briefly deformed, by a puff of air, for example, it vibrates, ringing at several different frequencies. The eigenfrequencies of these shape oscillations were determined by Lord Rayleigh,
\begin{equation}
  \sigma_T = \left(\frac{Tl(l-1)(l+2)}{\rho a^3}\right)^{1/2} \unit{rad \,s^{-1}},
\label{rayleigheq}
\end{equation}
where $T$ is the surface tension, $\rho$ is the density and $a$ is the radius of the spherical drop at rest \cite{rayleigh1879,lamb}.
By measuring $\sigma$, we can determine the surface tension of the liquid. Beaugnon et al. used diamagnetic levitation to measure $\sigma$ of the lowest order ($l=2$) mode of a diamagnetically levitated liquid droplet \cite{beaugnon01} and, recently, we used diamagnetic levitation to investigate dynamics of a spinning water droplet \cite{hill08}. A diamagnetically-levitated droplet is confined within a magnetogravitational potential trap \cite{berry97,simon00}. The trap acts as an additional cohesive force on the drop, perturbing its eigenfrequency spectrum. Beaugnon et al. observed the shift to higher frequency of the lowest-order $l=2$ mode, and determined an expression for the increase in terms of an enhanced effective surface tension \cite{beaugnon01}. However, their result cannot be generalized to the higher order modes. Here, we use diamagnetic levitation to measure the normal mode frequencies of a levitating drop, for $l\ge 2$. We derive an expression for the eigenfrequencies of a liquid droplet confined by the magnetogravitational potential trap and compare it with our measured frequencies. Our analysis points the way to achieving accurate measurements of surface tension using this non-contact technique. We also consider the effect on the eigenfrequencies of the shape of the potential trap and consider analogies with the vibrations of a model `star'.

\begin{figure}
 \includegraphics[width=85mm]{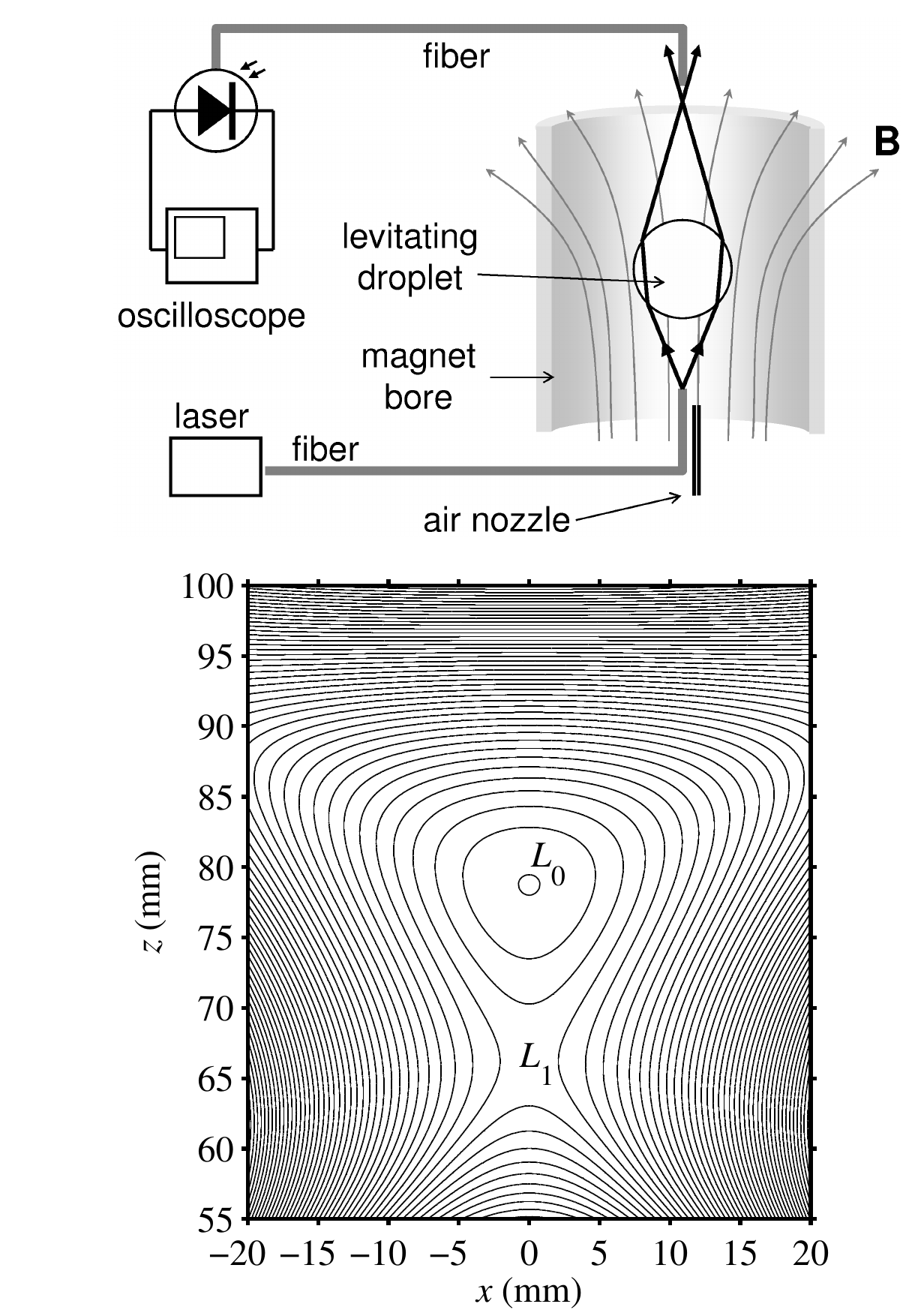}
 \caption{Bottom: Magnetogravitational potential $U(x,z)$ of a water droplet levitated by the magnetic field $B(x,z)$ of a vertical-bore superconducting solenoid magnet (contours at $\Delta U/g$ = 0.05~mm intervals); $x$ and $z$ are radial and vertical cylindrical coordinates, respectively, with origin at the geometric center of the solenoid. The field at the center of the solenoid is $B_0 = B(0,0) = 16.5\unit{T}$.  The field profile $B(x,z)$, used to construct the plot, was computed by numerical integration of the Biot Savart integral. The stable levitation point at a local minimum in the potential is labeled $L_0$. An unstable levitation point at a saddle-point in the potential is labeled $L_1$. Top: schematic of the experimental set-up. The magnetic field lines are shown in gray. Rays are shown as bold lines.}
\label{f1}
\end{figure}

\section{Experimental details}
\label{experimental_details}
We use a vertical-bore superconducting solenoid magnet with a room-temperature, 50~mm diameter bore to levitate droplets of water with radii $\sim 1 \unit{cm}$.  The droplets levitate approximately 80~mm above the geometric center of the solenoid, where the diamagnetic force, proportional to $B\nabla B$, is equal in magnitude to, and opposite in direction to, the gravitational force on (i.e. the weight of) the droplet \cite{berry97,simon00}. The magnetic field is $B\approx 12\unit{T}$ and the vertical field gradient is $\partial B/\partial z \approx 120\unit{Tm^{-1}}$ at the levitation  point.

\Fig{f1} shows a spatial map of the magnetogravitational potential
\begin{equation}
U(x,z) = gz - \frac{\chi B^2(x,z)}{2\rho \mu_0}
\label{magnetogravitationalpotential}
\end{equation}
of a unit mass of water in the potential trap, where $x$ and $z$ are radial and vertical cylindrical coordinates with origin at the geometric center of the solenoid coil, $\chi= -9\ten{-6}$ and $\rho=1\ten{3} \unit{kgm^{-3}}$ are the SI volume magnetic susceptibility and density of water respectively, and $g=9.8\unit{ms^{-2}}$. The center of the levitating droplet coincides with the position of the stable levitation point $L_0$ at the center of the trap, as shown. (There is a second point of unstable levitation $L_1$, at the saddle point in $U$, as shown in \fig{f1}. However, this point is not useful for these experiments, and so we do not discuss it further here.) The net force on a unit mass of water, $\gvec = -\nabla U$, is zero at the levitation point. The equilibrium shape of a liquid with no surface tension follows the contours of $U$, as demonstrated recently, using liquid H$_2$ close to the critical point \cite{lorin09}. For water droplets with $a\sim 1\unit{cm}$, however, the surface tension dominates the magnetic and gravitational forces on the drop, so that its equilibrium shape is nearly spherical.

The shape of the magnetogravitational potential trap can be altered by adjusting the current in the magnet solenoid coils. A convenient measure of the current is the magnetic field at the geometric center of the solenoid, $B_0 = B(0,0)$. The field $B(x,z)$ is everywhere proportional to $B_0$. To levitate water requires $B_0 \approx 16-17 \unit{T}$ using our magnet. \Fig{f1} shows the $U(x,z$) for $B_0 = 16.5\unit{T}$. The field profile $B(x,z)$ of the magnet, used to construct this plot, was computed by numerical integration of the Biot Savart integral, using a thin-shell approximation for the current density in the solenoid.

We expand the potential $U$ in a multipole expansion about $L_0$:
\begin{equation}
U(r,\theta) = \sum_{j\geq0} c_j(r) P_j(\cos\theta),
\end{equation}
where $r$ and $\theta$ are spherical coordinates with origin at $L_0$; $\theta$ is the polar angle (i.e. $r \sin \theta = x$). Only the derivative of $U$ normal to the droplet's surface (i.e. the radial component of the force) influences the eigenfrequencies of the normal modes,
\begin{equation}
\label{multipolederivative}
\gmag_r(r,\theta)=-\frac{\partial U(r,\theta)}{\partial r} = \sum_{j\geq0} c_j'(r) P_j(\cos\theta),
\end{equation}
where $c_j'=\partial c_j/\partial r$. By adjusting $B_0$, we reduce the quadrupole component until it is small compared to the spherically-symmetric component ($c_2' \ll c_0'$). The octopole harmonic ($c_3'$), which cannot be reduced this way, remains comparable to $c_0'$. All other harmonics are small compared to $c_0'$. We shall return to discuss these points further in \sct{nonspherical_potential}.

 \fig{f1}, top, shows a schematic diagram of the water droplet levitating in the vertical magnet bore.  The droplet at rest is close to spherical; we measure the ratio of the equatorial (horizontal) diameter to the polar (vertical) diameter to be $1.00\pm0.02$.  An optical fiber directs light from a HeNe laser at the droplet. The drop focuses the light onto the aperture of a second fiber, which transmits the light to a photodiode outside the magnet. The EMF of the photodiode is measured by a storage oscilloscope. A 1~mm-diameter nozzle, directed at the center of the underside of the droplet (see Fig. 1), is connected to a rubber bulb outside the magnet by a tube. When the bulb is struck on a hard surface, the resulting pulse of air from the nozzle excites several shape oscillation modes simultaneously, with amplitude $< 0.05 a$. Since the focal length of the drop depends on its shape, the intensity of the laser light falling on the photodiode oscillates as the drop vibrates. The temperature of the water was brought to $16^{\circ}$C in a water bath before it was injected into the potential trap, to match the ambient temperature in the magnet bore. At this temperature, the density of the liquid is $\rho = 999 \unit{kg m^{-3}}$, its surface tension is $T = 73.3 \unit{mNm^{-1}}$ and its kinematic viscosity is $\nu = 1.11 \ten{-6}\unit{m^2s^{-1}}$ \cite{lemmon}. The liquid was injected into the trap using a glass pipette. The volume of liquid injected was determined to better than $1\%$ uncertainty from the difference in the weight of the pipette before and after injection. The liquid was drawn out of the bore using a paper towel, by capillary action, after the experiment. By measuring the difference in weight between the wet and dry paper, we obtained a second measurement of the droplet volume. Using this simple and accurate technique, we were able to determine that the mass loss through evaporation of the drop, during the measurement period (approximately 30-60 minutes), was always less than $2\%$. Experiments were performed both in air and in dry nitrogen gas (by filling the bore with N$_2$ from a pressurized gas cylinder). The air experiments were performed at $B_0=16.2 \unit{T}$  and the nitrogen experiments at $16.5 \unit{T}$. The reason for the different $B_0$ may be explained by the oxygen content of the air, which, being paramagnetic, buoys up the droplet \cite{catherall05,catherall03} by an additional force $\gvec_{\mathrm{air}} = -\nabla U_{\mathrm{air}}$. Here, $U_{\mathrm{air}}=  \chi_{\mathrm{air}}B^2/(2\mu_0 \rho)$ and $\chi_{\mathrm{air}} = +3.7\ten{-7}$ is the volume magnetic susceptibility of air (S.I. units). The $c_2'$ component of $U + U_{\mathrm{air}}$ is minimized at $B_0=16.2 \unit{T}$. In nitrogen, the buoyancy force is negligible compared to the diamagnetic force, and $c_2'$ is minimized at $16.5 \unit{T}$ as shown in \fig{f1}.

\begin{figure}
\includegraphics[width=87mm]{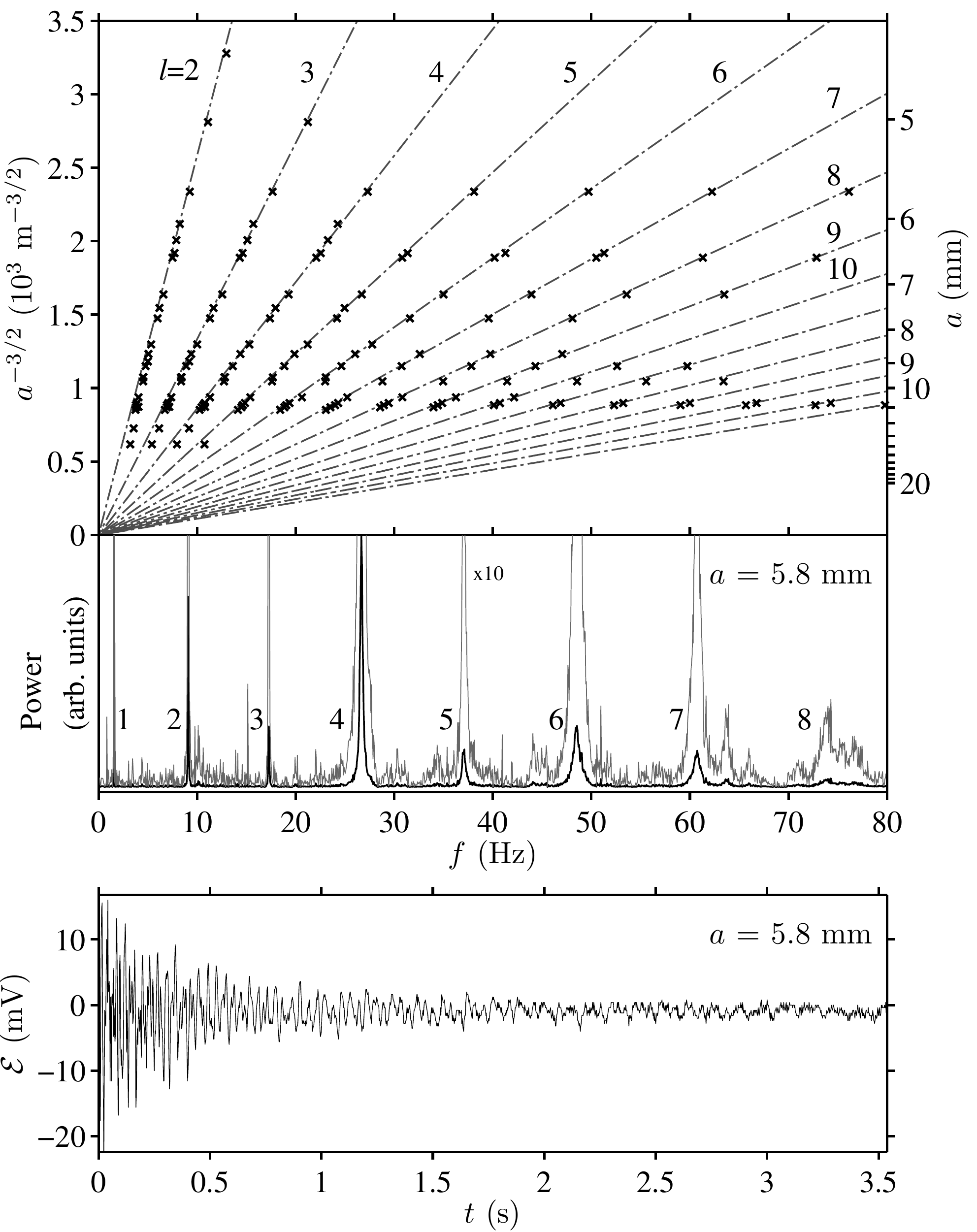}
 \caption
 {Bottom: photodiode voltage $\photovolt$ measuring the shape oscillations of an $a=5.8 \unit{mm}$ water droplet. Center: power spectrum of the oscillations of an $a=5.8 \unit{mm}$ water droplet, as a function of oscillation frequency $f=\omega/2\pi$. Top: measured frequencies $f$ of the oscillation modes of water droplets with radii between $a=4.5\unit{mm}$ and $14\unit{mm}$, up to $l=16$. The broken lines show the Rayleigh frequency spectrum for a water drop, according to \eqn{rayleigheq}. The \emph{measured} frequencies (crosses) are slightly higher, by $\sim 1 \unit{Hz}$, due to the effect of the magnetogravitational potential trap.}
 \label{f2}
\end{figure}
\section{Results}
\Fig{f2}, bottom, shows the oscillations in the photodiode EMF $\photovolt(t)$ developed by the light refracted through an $a=6\unit{mm}$ droplet; they decay exponentially with a time constant $\tau \sim 0.1-10\unit{s}$, dependent on $a$ and $l$, due to the viscosity of the water, i.e. $\photovolt(t) = \sum_l \photovolt_l \exp(i\omega_l t - t/\tau_l)$. The power spectrum of these oscillations is shown in \Fig{f2}, center. Several peaks are evident in this spectrum, corresponding to the $l=2$ to 7 Rayleigh modes of the drop. Peak 1 is due to a small oscillation of the droplet's center of mass of about $L_0$ (amplitude $\sim 1\unit{mm}$). The frequency of this peak is independent of $a$.
The upper panel of \fig{f2} shows the measured peak positions for different drops with radii between $a=4.5\unit{mm}$ and 14~mm; the broken lines show the Rayleigh frequency spectrum for a water drop, according to \eqn{rayleigheq}. Note that the measured frequencies are slightly \emph{higher}, by $\sim 1\unit{Hz}$, due to the effect of the magnetogravitational potential trap. We shall calculate this frequency shift in the following sections of the paper. Although small, it is important to be able to account for this shift, if the technique is to be used to obtain accurate measurements of surface tension, for example.

An estimate of the effect of viscosity on the frequency of oscillation can be obtained by calculating the ratio of the magnitude of viscous stresses to surface tension forces, $f\nu\rho/(TD^{-1}) = \OhnesorgeNumber[2l(l-1)(l+2)]^{1/2}/\pi$, where $D = 2a$, $f=2\pi\omega$ and $\OhnesorgeNumber$ is the Ohnesorge number \cite{lefebvre}. This ratio is much smaller than unity for the droplet sizes used in these experiments. This indicates that we can neglect the influence of viscosity on the oscillation frequencies as a small effect. For example, we expect the viscosity of the water to marginally lower the frequency $\omega$ of the $l=2$ mode of an $a=5\unit{mm}$ droplet by $(5\pm 1)\ten{-4}\%$, but this is a small reduction compared to the frequency increase resulting from the trapping potential. (We obtained these estimates by solving numerically the Chandrasekhar equation for the eigenfrequencies of a viscous drop \cite{chandra59, tang74}). The $\OhnesorgeNumber$ number indicates that viscosity has a significant effect on the frequency of modes $l<20$ only for water drops smaller than $a \sim 10\unit{\mu m}$. The shape of the peaks in the power spectrum agrees well with the Lorentzian shape expected for an exponentially decaying oscillation, with half-width (HWHM) $\Delta \omega =1/\tau \approx \nu a^{-2}(l-1)(2l+1)$ given by the Chandrasekhar equation  \cite{chandra59, tang74}. (Since the power spectrum is the square of the magnitude of the Fourier transform of the oscillations, we compare the shape of the peak with the \emph{square} of the Lorentzian function).

\section{Spherical potential approximation}
We now consider the effect of a spherically-symmetric magnetogravitational potential well on the eigenfrequencies of the drop. We shall discuss the effects of additional harmonics in \sct{nonspherical_potential}. Our derivation follows Lamb's derivation of the Rayleigh frequencies \cite{lamb} closely, with the addition of the force $\Gamma_r$ on the droplet's surface due to the gradient of the magnetogravitational potential at the surface. We write the shape of the $l$th mode of a drop oscillating with frequency $\omega$, for oscillations with small amplitude $\epsilon$, as
\begin{equation}
r= R(\theta, t) = a + \zeta = a + \epsilon P_l(\cos\theta)\sin \omega t
\end{equation}
where $P_l$ is a Legendre polynomial of degree $l\ge1$ ($l=1$ corresponds to an oscillation of the droplet's center of mass in the potential trap).

The pressure equation at the surface of the droplet (to first order) is \cite{lamb}
\begin{equation}
   \left.\frac{\partial \phi}{\partial t}\right|_{r=a} = U(R) + \frac{p(R)}{\rho}+ F(t),
\label{pressureeqn}
\end{equation}
where $F(t)$ is an arbitrary function of $t$ only and $\phi$ is the velocity potential \cite{lamb}
\begin{equation}
   \phi(r,\theta) = -\frac{r^l}{la^{l-1}}\omega \epsilon P_l(\cos\theta)\cos \omega t.
\label{e3}
\end{equation}
The pressure difference across the surface resulting from the surface tension is \cite{lamb}
\begin{equation}
p(R) = T\left(\frac{2}{a} + \frac{\zeta(l-1)(l+2)}{a^2} \right).
\label{lambpressure}
\end{equation}
The magnetogravitational potential $U$ at the surface of the drop is (to first order)
\begin{subequations}
\begin{align}
U(R(\theta,t)) &= U(a) - \gmag_r(a,\theta)(R-a)\label{potlexp2} \\
               &= U(a) - \epsilon \gmag_r(a,\theta)P_l(\cos\theta) \sin(\omega t)\label{potlexp},
\end{align}
\end{subequations}
where $\gmag_r(a,\theta) = -c_0'(a)$ for a spherically symmetric well (see \eqn{multipolederivative}). Inserting Eqns.~\ref{potlexp}, \ref{lambpressure} and \ref{e3} into \eqn{pressureeqn}, we obtain
\begin{equation}
\omega^2 = \sigma_T^2 + \sigma_0^2,
\label{measuredspherical}
\end{equation}
where
\begin{equation}
\label{sigma_gmag}
\sigma_0^2 = \frac{c_0'(a)l}{a}
\end{equation}
is the oscillation frequency (squared) of a hypothetical drop with $T=0$, held together by the magnetogravitational trap alone. The fact that the square of the \emph{measured} frequency $\omega^2$ is a simple sum of the square of the Rayleigh frequency $\sigma_T^2$ and $\sigma_0^2$ is due to the fact that
$l$ remains a good eigennumber for oscillations in a \emph{spherically-symmetric} potential well. In \sct{nonspherical_potential}, we will consider the effect of a non-spherical magnetogravitational well. In a non-spherical well, $l$ is not a good eigennumber in general, but for small deviations from spherical, we can use perturbation theory to obtain corrections to \eqn{measuredspherical}.

We can obtain an experimental measurement of $c_0'$ by examining the difference between the measured frequencies $\omega$ of any two of the modes $l,n\ge2$:
\begin{equation}
h_l\sigma_{0,n}^2-h_n\sigma_{0,l}^2 = h_l\omega_n^2 - h_n\omega_l^2,
\end{equation}
where $h_l=l(l-1)(l+2)$. Dividing by $(nh_l - lh_n)/a$, we obtain an experimental measurement of $c_0'(r)$ from the oscillations of a drop that has radius $a=r$ at rest:
\begin{equation}
\label{gmagmeasure}
c_0'(r) = r\frac{h_l\omega_n^2 - h_n\omega_l^2}
                 {nh_l-lh_n}.
\end{equation}
By using two measured frequencies, $\omega_l$ and $\omega_n$, rather than a single frequency, we obtain a measurement of $c_0'$ independent of the surface tension.
\begin{figure}
\includegraphics[width=82mm]{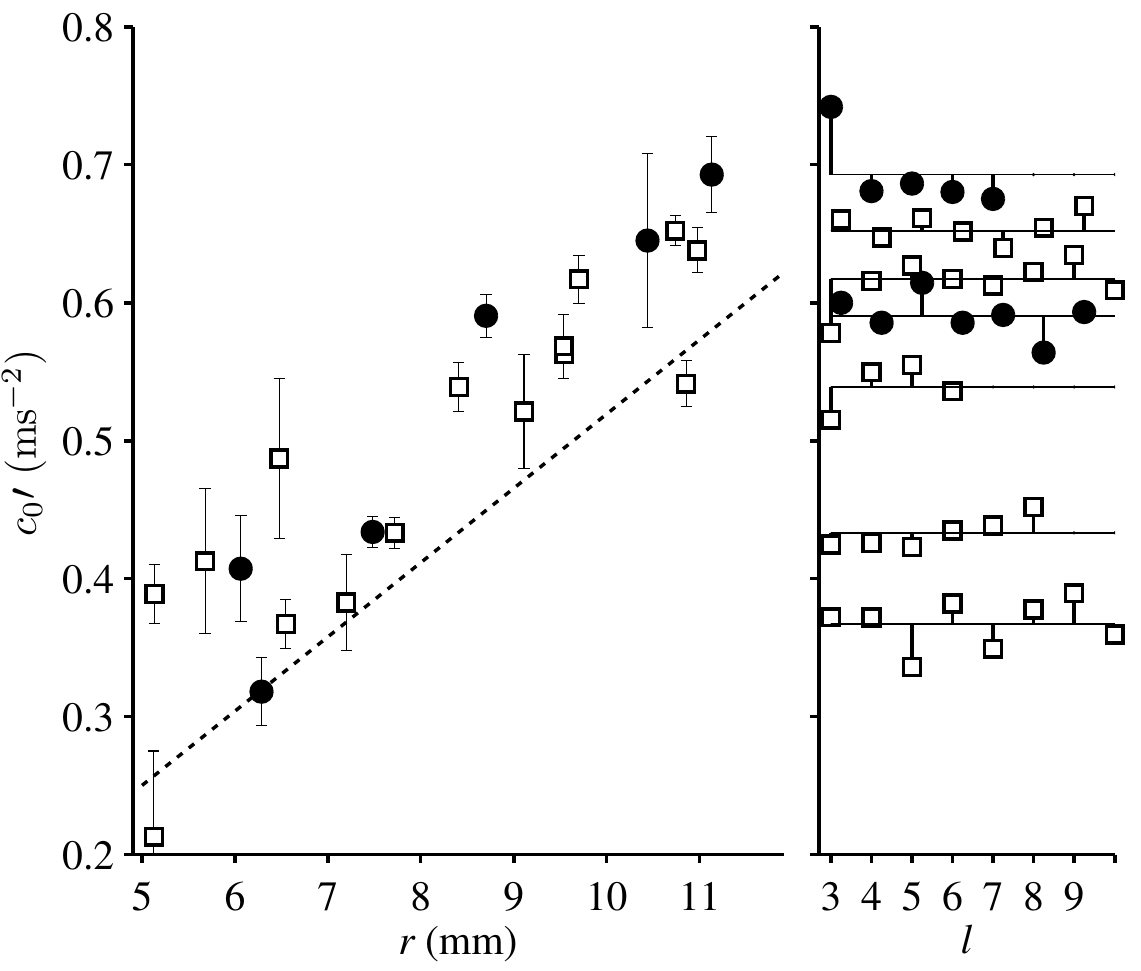}
 \caption
 {
    Left panel: Gradient of the magnetogravitational potential trap $c_0'$ obtained from the measured frequencies of two modes, $l=2$ and $l>2$, of a drop with radius $a=r$ at rest. Error bars show the standard error. Filled circles: data obtained in nitrogen atmosphere. Open squares: data obtained in air atmosphere. Dotted line: $c_0'$ computed from the solenoid geometry. Right panel: dependence of the experimentally-obtained $c_0'$ on the mode number with $l>2$ (see text).
 }
 \label{f3}
\end{figure}
\Fig{f3} (left panel) shows the values of $c_0'$ that we obtain from the above `two-frequency' method using the lowest order mode $n=2$ and a second mode $l>2$. We use the $n=2$ mode since the corresponding peak in the power spectrum is always clearly resolved. We plot the mean of the values obtained for each mode $l>2$. Error bars (standard error) indicate the variation in the measured $c_0'$ obtained from different modes $l$. Filled circles and open squares show data obtained from a drop in a nitrogen atmosphere at $B_0=16.5\unit{T}$, and in air at $B_0=16.2\unit{T}$, respectively. The broken line on \fig{f3} shows the value of $c_0'(r)$ computed from the potential $U$ shown in \fig{f1} (i.e. from the geometry of the solenoid and the current $B_0$). This line is in reasonably good agreement with the experimentally measured values of $c_0'$, although the data points fall at slightly higher values. As an additional check of our method, we plot the dependence of the measured $c_0'$ on $l$ (right hand panel). If the trap-induced frequency enhancement that we measure experimentally is accurately given by \eqn{sigma_gmag}, then the $c_0'$ values that we determine by this method should exhibit no dependence on $l$. Although there is some scatter in the data due to experimental error, there is no clear dependence on $l$.

\section{Non-spherical potential}
\label{nonspherical_potential}
In the previous section, we assumed that the potential well was spherically symmetric around $L_0$. However, whilst we have chosen the field $B_0$ to minimize the quadrupole component of the trap, the octopole component remains significant, as can be seen clearly in \Fig{f1}. Thus it initially appears surprising that we can treat the well \emph{as if} it were spherical, in order to calculate its effect on the vibrations of the droplet's shape. (Had we determined an unphysical dependence of our measured $c_0'$ on $l$ in the previous section, it would also have followed that the spherical-well approximation was inadequate.) We now consider the effect on the eigenfrequencies of the drop of additional harmonic components $c_j'$ of the potential trap, and consider why, if the trap has a significant octopole component, the spherical well approximation is so effective at reproducing the measured frequencies of the droplet.

\begin{figure}
\includegraphics[width=82mm]{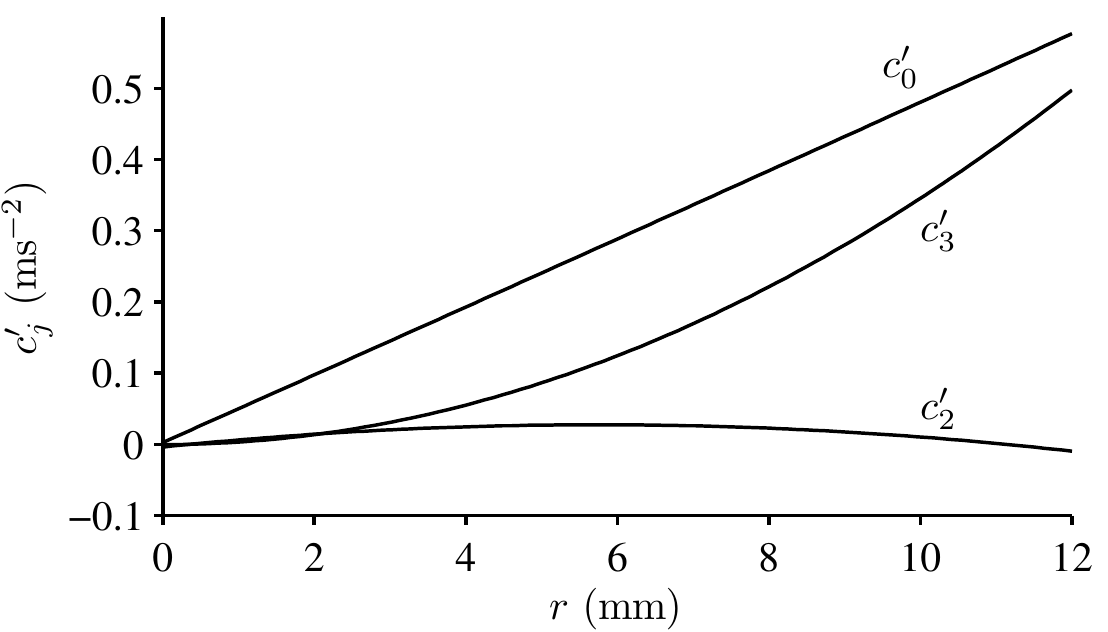}
 \caption
 {
     Values of the $j=0$ (spherically-symmetric) and $j=2,3$ (quadrupole and octopole) coefficients $c_j'$ in a multipole expansion of the potential well gradient, \eqn{multipolederivative}, at $B_0 = 16.5 \unit{T}$ (in nitrogen atmosphere). Note, $c_1' = 0$ (see text). The values were determined from the magnetogravitational potential (\eqn{magnetogravitationalpotential}) shown in \fig{f1}. The $c_0'$ line is the same as that shown by the dotted line on \fig{f3}.
 }
 \label{figcoj}
\end{figure}

\begin{figure}
\includegraphics[width=82mm]{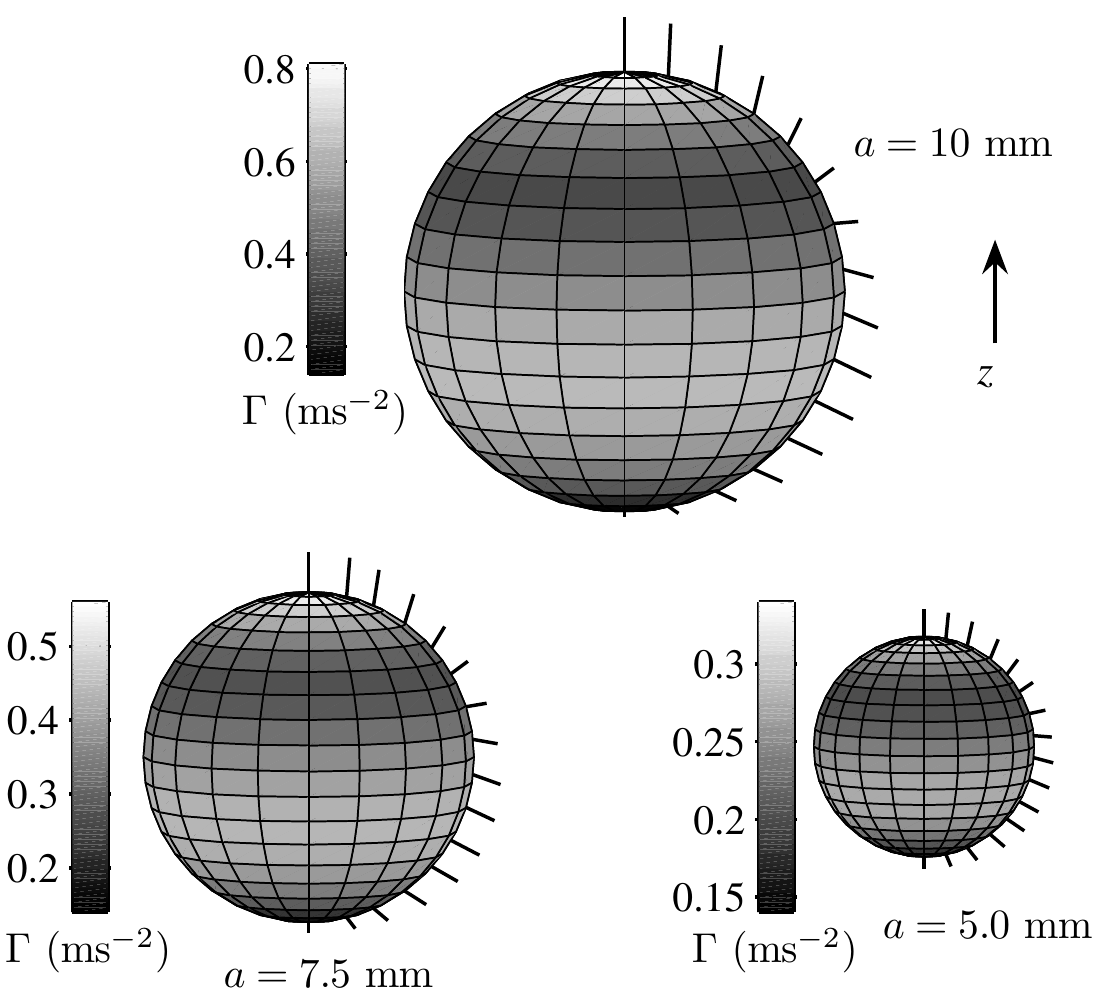}
 \caption
 {
   The three spheres show the computed magnitude of the vector $\gvec=-\nabla U$ on the surface of three droplets, radius $a=5.0\unit{mm}$, $7.5\unit{mm}$ and $10.0\unit{mm}$ at $B_0 = 16.5 \unit{T}$ (in nitrogen atmosphere); the length and direction of the radiating lines show the direction and magnitude of the (inward pointing) $\gvec$ at the surface; $\gvec$ was determined from the magnetogravitational potential (\eqn{magnetogravitationalpotential}) shown in \fig{f1}. The variation in $|\gvec|$ over the surface of the droplet is due to the octopole component of the potential trap.
 }
 \label{figspheres}
\end{figure}

We first discuss how close to spherical it is possible to make the potential trap; that is, how small the coefficients $c_j'$, $j>0$ can be made. Lorin and Mailfert have considered the relationship between the coil geometry (the distribution of the current density) and the shape of the potential trap \cite{lorin08}. Although we cannot alter the coil geometry, we can alter the coefficients $c_j'$ by adjusting the current flowing in the solenoid. The magnitude of the quadrupole component of the radial force $|c_2'|$ is smallest (for our magnet) at $B_0=16.2\unit{T}$ in air, and at $B_0=16.5\unit{T}$ in nitrogen gas; at this field, $|c_2'| \ll |c_0'|$ and only weakly dependent on $r$. Increasing $B_0$ beyond this value gives a positive $c_2'$, increasing with $r$. Decreasing $B_0$ below this value gives a negative $c_2'$, becoming more negative with increasing $r$. The gradient of the octopole component $c_3'$, however, is of the same order as $c_0'$ and increases as $r^2$. This component is a feature of the magnetogravitational potential trap generated by a solenoid. It can be reduced slightly, by increasing $B_0$, but only at the expense of significantly increasing $|c_2'|$. Note that the dipole component of the trap $c_1$ is necessarily zero (hence $c_1'=0$ too), since the net vertical force on the droplet $F_z$ must be zero for levitation. Since $F_z$ is proportional to the difference between the mean surface potentials of its upper and lower hemispheres, i.e. $F_z \propto \int_0^1 |w| U(a,w) dw - \int_{-1}^0 |w| U(a,w) dw \propto \sum_j c_j \int_{-1}^1 w P_j(w) dw \propto c_1$, it follows that $c_1=0$, where $w=\cos\theta(=P_1)$. The harmonic components $j>3$, are small compared to both $c_0'$ and $c_3'$, in our system. \Fig{figcoj} shows the values of the $j=0$ and $j=2,3$ (quadrupole and octopole) coefficients $c_j'$ at $B_0 = 16.5 \unit{T}$ (for nitrogen atmosphere) computed from the solenoid geometry. The spheres in \fig{figspheres} show the computed magnitude of $\gvec = -\nabla U$ at the surface of three droplets, radius $a=5.0\unit{mm}$, $7.5\unit{mm}$ and $10.0\unit{mm}$. Radiating lines indicate the direction and magnitude of $\gvec$, which is directed toward the interior of the droplet. The variation in $|\gvec|$ over the surface of the droplet is due to the octopole component of the potential trap.

\begin{figure}
\includegraphics[width=87mm]{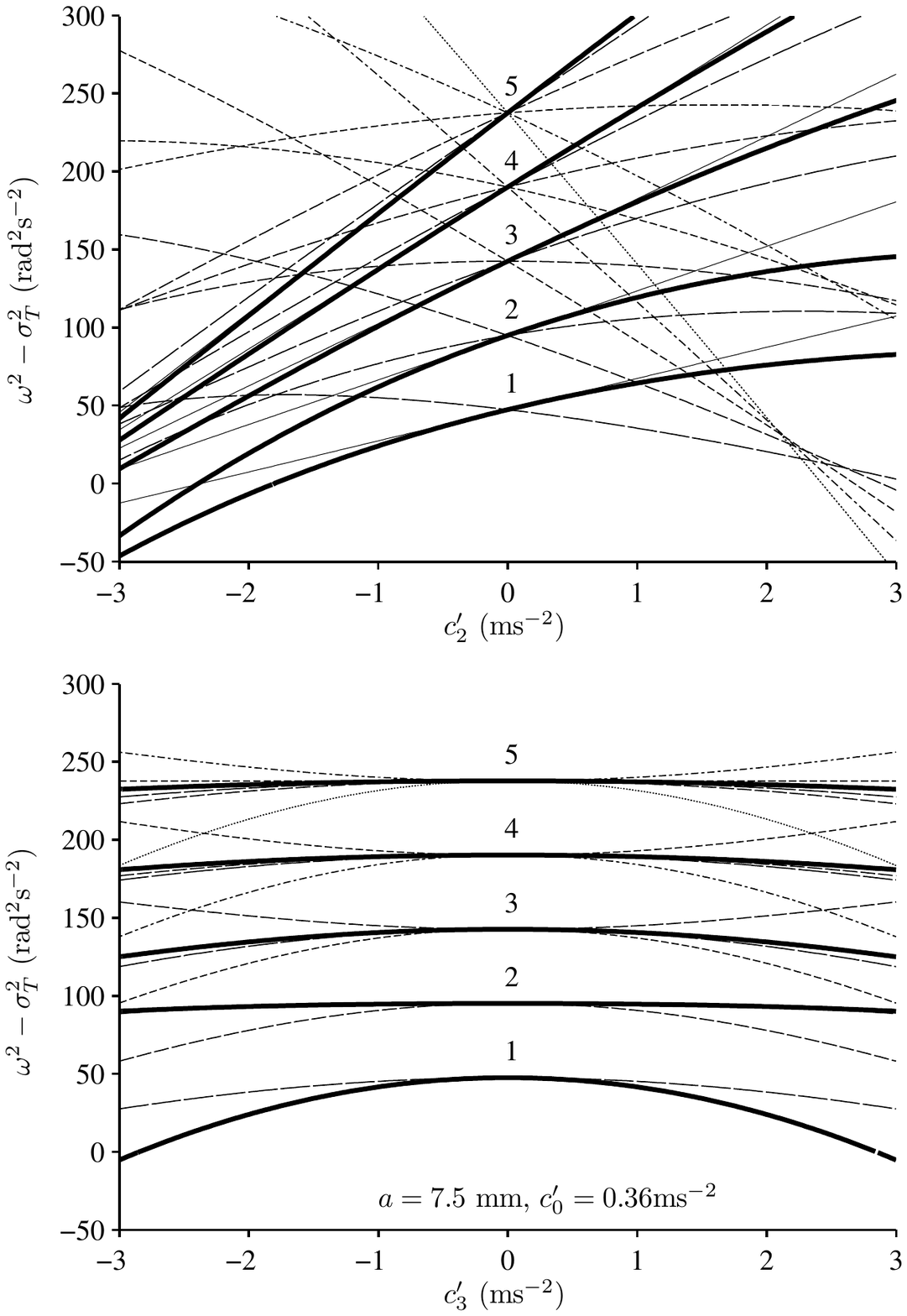}
 \caption{Top: Effect on the eigenfrequencies of an $a = 7.5 \unit{mm}$ water droplet of adding a quadrupole harmonic $c_2'$ to the shape of the magnetogravitational potential well ($c_0' = 0.36\unit{ms^{-2}}$, all other $c_j' = 0$). The thick black lines show the eigenfrequencies (squared) of the $m=0$ modes (i.e. the eigenvalues $\omega^2$ of \eqn{eigenvalueproblem}) corresponding to $l=1-5$; the eigenvalues of \eqn{eigenvalueproblem} were computed using MATLAB. (Note that we have plotted $\omega^2-\sigma_T^2$ for clarity). The thin lines show the first-order approximation, \eqn{firstorderquadrupole}, departing from the numerically-computed frequencies at $|c_2'|\gtrsim 1$. Long-dash,  medium-dash, short-dash, dot-dash and dotted lines show the frequencies of the $|m|=1-5$ modes, respectively (the treatment of the $m\neq0$ modes is summarized in the Appendix). Bottom: Effect on the eigenfrequencies of an $a = 7.5\unit{mm}$ water droplet of adding a harmonic $c_3'$ to the shape of the well ($c_0' = 0.36\unit{ms^{-2}}$, all other $c_j' = 0$).
 }
 \label{eigenvalues}
\end{figure}

We now consider the effect on the droplet's eigenfrequencies of adding a harmonic component $c_j'$, $j\ge2$ to the potential trap. The analysis proceeds as above, however, we must now include higher order harmonics in the eigenfunction of the shape oscillation
\begin{equation}
r= R(\theta, t) = a + \epsilon \sin \omega t \sum_{l\geq1} b_{l}P_l(\cos\theta),
\label{shapegeneral}
\end{equation}
since $l$ is not, in general, a good eigennumber in a non-spherical potential. In principle, we should decompose the shape into spherical harmonics $Y_l^m$, since the degeneracy in $m$ is also lifted in a non-spherical potential (see \fig{eigenvalues}). However, our method of inducing shape oscillations in the droplet tends to excite only the axisymmetric shapes (i.e. with $m=0$), since the air jet is aligned along the solenoid axis. For this reason, we derive here the frequencies of the $m=0$ oscillations only (\eqn{shapegeneral}), which are sufficient to interpret the experimental results. We summarize the treatment of the general case $|m|\le l$ in the Appendix.

The velocity potential is
\begin{equation}
   \phi(r,\theta) = -\epsilon\omega\cos \omega t \sum_{l\geq1}b_{l}r^ll^{-1}a^{-l+1} P_l(\cos\theta).
\label{phigeneral}
\end{equation}
The magnetogravitational potential $U$ at the surface of the drop is (see \eqn{potlexp2})
\begin{equation}
   U(R) = U(a) - \epsilon \gmag_r(a,\theta) \sin(\omega t)\sum_{l\ge1} b_l P_l(\cos\theta)
\label{potlexp3},
\end{equation}
where $-\gmag_r(a,\theta) = c_0'(a) + c_j'(a)P_j(\cos\theta)$ in this case (see \eqn{multipolederivative}).
Inserting \eqn{potlexp3}, \eqn{shapegeneral} and \eqn{phigeneral} into \eqn{pressureeqn}, and equating the time-varying terms, we obtain
\begin{equation}
a\omega^2 \sum_{l\geq1}\frac{b_lP_l}{l} = \sum_{l\geq1}\left[(c_0'(a)+c_j'(a)P_j) +\frac{T}{\rho a^2}(l-1)(l+2)\right]b_lP_l.
\end{equation}
The product $P_lP_j$ appearing on the RHS of this equation can be expanded as a sum of Legendre polynomials \cite{adams1878}, which, for our purposes, is most conveniently written $P_lP_j=\sum_{p=|j-l|}^{j+l} Q(l,j,[j+l-p]/2)P_{p}$, in which \cite{adams1878}
\begin{equation}
  Q(l,j,s)=\frac{A(l-s)A(s)A(j-s)}{A(j+l-s)}\left(\frac{2j+2l-4s+1}{2j+2l-2s+1}\right)
\end{equation}
for integer $s$ and we define $Q(l,j,s)=0$ for half-integer $s$; $A(n) = 1\times3\times5\times\ldots\times(2n-1)/n!$. Equating the coefficients of $P_l$, we obtain
\begin{align}
\omega^2b_l &=  \frac{c_0'(a)l}{a}b_l + \frac{T}{\rho a^3}l(l-1)(l+2)b_l\nonumber \\
&  + \frac{c_j'(a)l}{a}\sum_{p=|j-l|}^{j+l} b_{p}Q(p,j,[j+p-l]/2).
\label{eigenvalueproblem}
\end{align}
This equation has the form of an eigenvalue problem $\omega^2b_l = H_{l\lambda}b_\lambda = (H^{(0)}_{l\lambda} + V_{l\lambda})b_\lambda$ (using the summation convention, $l\geq1$), where $H^{(0)}$, representing the first two terms on the RHS of \eqn{eigenvalueproblem} is a diagonal matrix and $V$, representing the third term on the RHS is not, in general, diagonal. Treating $V$ as a perturbation, the first order correction $(\sigma_j^2)^{(1)} =b_l^{(0)}V_{l\lambda}b_\lambda^{(0)}$ due to a harmonic component $j$ can be computed analytically to obtain the eigenfrequencies $\omega^2 = \sigma_T^2 + \sigma_0^2 + (\sigma_j^2)^{(1)} + \ldots$. For a quadrupole harmonic $j=2$ we obtain
\begin{equation}
  (\sigma_j^2)^{(1)}=\frac{c_2'(a)}{a}\frac{l^2(l+1)}{(2l+3)(2l-1)}.
  \label{firstorderquadrupole}
\end{equation}
For an octopole harmonic $j=3$, and all odd $j$, $(\sigma_j^2)^{(1)}=0$. This explains why our spherical-well approximation works so well in predicting the eigenfrequency spectrum of the droplet: we have minimized $c_2'$ by careful adjustment of $B_0$ and we expect, from the above analysis, that the effect of the octopole harmonic $c_3'$ (which is comparable to the spherically symmetric component $c_0'$) on the measured eigenfrequencies in our experiment, to be minimal. \Fig{eigenvalues} shows the eigenfrequencies determined by using a computer to solve the eigenvalue problem numerically (i.e. beyond first order). The eigenfrequencies of an $a=7.5 \unit{mm}$ droplet, for various $c_2'$ and $c_3'$, are shown (thick lines), along with the first-order result \eqn{firstorderquadrupole} for comparison (thin lines). In addition, we plot the eigenfrequencies of modes with $m\neq0$ (broken lines); the calculation of the frequency of these modes is outlined in the Appendix. The lower plot of \fig{eigenvalues} shows that the effect of the octopole harmonic $c_3'$ on the measured eigenfrequencies is minimal for all $m$ for $|c_3'|\lesssim 1$. For a quadrupole harmonic $c_2'$, the eigenfrequencies depend strongly on $m$, as shown in the upper plot of \fig{eigenvalues}. However, since we have reduced $|c_2'|$ to smaller than $\approx 0.05$ as described above, and since our method of exciting the oscillations tends to excite only the $m=0$ modes, this does not alter our explanation of why the spherical-well approximation works so effectively.

\section{Discussion and Conclusion}
Although the forces at the droplet's surface are dominated by the surface tension, we can clearly observe the influence of the magnetogravitational potential trap on the eigenfrequencies. We have demonstrated that the effect of the trap on the eigenfrequencies closely approximates that of a spherically-symmetric potential well. Although the potential well has a significant octopole component in additional to the spherically-symmetric component, we have calculated that the effect of the octopole component on the eigenfrequencies is small and verified this experimentally. The small difference between the values of $c_0'$ calculated from the solenoid geometry and the measurement of $c_0'$ from the measured droplet oscillations, may be due to a small error in the calculated value: the calculation of the field profile $B(x,z)$, performed by numerical integration of the Biot-Savart equation, is based on a thin-shell approximation of the current density in the solenoid coils. We note that the discrepancy cannot be explained by the octopole component of the trap, since this only marginally reduces the eigenfrequencies (\fig{eigenvalues}), which would only \emph{reduce} the measured value of $c_0'$ slightly. Experiments have shown that nonlinear effects become significant for oscillation amplitudes greater than approximately $0.1a$ \cite{becker91}. In our experiments, however, the oscillation amplitude is smaller than this, and we do not observe these non-linear effects.

We have shown how to minimize the quadrupole harmonic of the potential trap, by adjusting the solenoid current. This allowed us to use the relatively simple formula \eqn{sigma_gmag} to calculate the effect of the trap on the eigenfrequencies. This opens up the possibility of using diamagnetic levitation to accurately determine the surface tension of diamagnetic liquids (for example, water, and many water-based and organic solutions): subtracting the contribution of the trap to the eigenfrequencies reveals the Rayleigh-frequency spectrum, from which the surface tension can be obtained directly.
A contactless measurement technique has many advantages, for example, the ability to measure highly reactive liquids and to achieve significant supercooling of the liquid, as demonstrated in experiments on electromagnetically levitated liquid metals \cite{egry95}. Acoustic levitation and the oscillating drop method can be used to measure surface properties of small drops (radius $<2\unit{mm}$) of water and organic liquids (e.g. \cite{trinh88, tian97}). Suspension of small drops in air-flow is also possible \cite{perez99}. However, since the equilibrium droplet shape is distorted significantly from spherical in both of these techniques, it is necessary to make accurate measurements of the equilibrium shape in order to correct for the distortion, which introduces additional significant experimental uncertainty \cite{trinh86,trinh88,tian97,perez99}. Drops can be bounced on a solid surface or a vibrating liquid bath, the drop and surface being separated by a thin layer of air \cite{couder05,gilet07}. In this case, the eigenfrequency spectrum deviates from the Rayleigh spectrum due to the periodic forcing by the oscillating surface \cite{dorbolo08,courty06}. An expression, similar to Rayleigh's, for the eigenfrequencies of vibrating drops in continuous point-contact with a surface, has been determined \cite{courty06} but there remains some uncertainty over the spectrum of bouncing drops \cite{courty06}. Although free-fall can be used to obtain the Rayleigh frequencies directly (for example, in a drop-tower \cite{matsumoto02}), we have shown how diamagnetic levitation could offer an alternative, accurate method of measuring these frequencies; the levitated droplets are near-spherical at rest and the eigenfrequency spectrum is very close to Rayleigh's, as shown in \fig{f2}. The relatively small shift to higher frequency due to the potential trap can be obtained using a simple spherical-well approximation, \eqn{sigma_gmag}. Diamagnetic levitation can levitate cm-size drops (up to $\approx 3$ cm-diameter in our magnet) \cite{beaugnon01,hill08,liu10} enabling the droplet volume to be measured easily and precisely. Currently, there is interest in the temperature dependence of the surface tension of supercooled water \cite{lu06}. The lowest-temperature achieved so far, which used a contact technique, is 245~K \cite{floriano90}; nucleation sites on the container walls trigger freezing before colder temperatures can be reached. We propose that diamagnetic levitation and the oscillating drop technique could be used to obtain measurements of surface tension at temperatures less than 245~K.

It is interesting to consider an analogue between the oscillations of the levitating droplet and that of an object in a gravitational field. We can consider the magnetogravitational force on a unit mass of water, $\gvec = -\nabla U$,
as being an effective gravitational field acting on the liquid. Reid has obtained an expression for the eigenfrequencies of a gravitating body composed of a solid spherical core, radius $R_1$ and density $\rho_1$ covered by an inviscid liquid `mantle' of radius $R_2$, density $\rho_2$ \cite{reid59} (see also Ref. \cite{dolginov84}):
\begin{equation}
\label{reid}
\sigma^2 = \frac{4}{3}\pi G \bar{\rho}l(l+1)\frac{1-\eta^{2l+1}}{1+l(1+\eta^{2l+1})}\left(1-\frac{3}{2l+1}\frac{\rho_2}{\bar{\rho}}\right),
\end{equation}
where $\eta=R_1/R_2$, $\bar{\rho}=\eta^3\rho_1+(1-\eta^3)\rho_2$ is the mean density and $G$ is the gravitational constant. Interestingly, Reid's expression becomes equivalent to \eqn{sigma_gmag} if we let $\rho_2\rightarrow 0$ and $R_1\rightarrow 0$, since this allows the surface effective gravity $\gmag(a)$ to be equated with a fictitious point mass $m=(4/3)\pi a^3 \bar{\rho} = \gmag(a)a^2/G \sim 1\ten{8}\unit{kg}$ (approximately the mass of 1~mm$^3$ of nuclear-density material). In this sense, we can view our magnetically-levitated droplet as a `toy model' of an incompressible gravitating body (possessing surface tension), composed of a small-diameter, massive core (the fictitious mass), and a low density outer region (the water). Note that this result differs from that of a \emph{uniform-density} self-gravitating body, $\sigma_G = [(8/3) \pi G \rho
l(l-1)/(2l+1)]^{1/2}$ \cite{kelvin1863}. Bastrukov has recently obtained a similar spectrum for a self-gravitating liquid droplet with radially-varying density distribution $\rho(r<a) \propto 1/r$, having a singular density at the center \cite{bastrukov09}.

By adjusting the current in the magnet (i.e. $B_0$) we can change the quadrupole ($c_2'$) component of the trap. Although we have sought to minimize this component in this paper, it would be interesting to investigate the effect of this quadrupole component on the oscillation frequencies experimentally and compare it with the result we derive above. Unlike the odd harmonics, the even order harmonics have a significant effect on the eigenfrequencies to first order. Using the analogy with a model `star', adding a quadrupole component to the effective gravity $\gvec$ is equivalent to deforming the shape of the core mass from spherical to oblate, perhaps due to rotation, for example.

This project is supported by a Basic Technology Grant from EPSRC, UK; Grant Nos. GR/S83005/01 and EP/G037647/1.

\section{Appendix}
We consider the effect of a trap component $c_j'$ on an arbitrary shape oscillation
\begin{equation}
r= R(\theta, \Phi, t) = a + \epsilon \sin \omega t \sum_{l\geq1} \sum_{m=-l}^{l}b_{l}^m Y_l^m(\theta,\Phi).
\label{shapegeneralA}
\end{equation}
The corresponding velocity potential is
\begin{equation}
   \phi(r,\theta,\Phi) = -\epsilon\omega\cos \omega t \sum_{l\geq1} \sum_{m=-l}^{l} b_{l}^m r^ll^{-1}a^{-l+1} Y_l^m(\theta,\Phi).
\label{phigeneralA}
\end{equation}
Inserting \eqn{shapegeneralA} and \eqn{phigeneralA} into \eqn{pressureeqn}, and equating the time-varying terms, we obtain
\begin{equation}
a\omega^2 \sum_{l\geq1}\sum_{m=-l}^{l}\frac{b_l^m Y_l^m}{l} = \sum_{l\geq1}\sum_{m=-l}^{l}\left[(c_0'+c_j'P_j) +\frac{T}{\rho a^2}(l-1)(l+2)\right]b_l^m Y_l^m.
\end{equation}
The product $Y_l^m P_j = N_l^m e^{im\Phi}P_l^m P_j^0$ can be expanded as $N_l^m e^{im\Phi}\sum_{p=|j-l|}^{j+l} Q(0,j;m,l;p)P_p^m$, where the $Q$ in this case are Gaunt coefficients, as defined in Ref. \cite{cruzan62}, $P_l^m$ are associated Legendre functions and $N_l^m$ are the corresponding normalization factors. Equating the coefficients of $Y_l^m$, we obtain $2l+1$ eigenvalue problems, i.e. one for each $m$, similar to \eqn{eigenvalueproblem}.




\end{document}